\documentclass[12pt,journal,onecolumn,draftclsnofoot,]{IEEEtran}
\usepackage{amsmath}
\usepackage{amssymb}
\usepackage{graphicx}
\usepackage{color}
\usepackage{cite}
\usepackage{array}
\renewcommand\IEEEkeywordsname{Index Terms}
\begin{document}

\title{Multicarrier Chirp-Division Multiplexing for Wireless Communications} \vspace{-8pt}
\author{\IEEEauthorblockN{Song-Wen Huang and Dimitris A. Pados}

\thanks{Song-Wen Huang is are with the Department of Computer and Electrical Engineering and Computer Science, Florida Atlantic University, Boca Raton, FL 33431, USA. Dimitris A. Pados is with the I-SENSE, Department of Computer and Electrical Engineering and Computer Science, Florida Atlantic University, Boca Raton, FL 33431. (e-mail: songwenh@buffalo.edu; dpados@fau.edu)}
\vspace{-15pt}}

\maketitle

\begin{abstract}
We propose a multicarrier chirp-division multiplexing (MCDM) system in which orthogonal chirp waveforms are utilized as frequency subcarriers. Orthogonal characteristics of chirp subcarriers are analyzed in respect to cross-correlation coefficients among subcarriers. Moreover, orthogonal chirp transform (OCT) is developed to implement MCDM systems. In addition, we design a low complexity receiver, including packet synchronization, carrier frequency offset compensation, channel estimation, and symbol detection. Proposed MCDM systems have both advantages of chirp waveforms and multicarrier architectures. Computational complexity of our proposed detector is discussed in detail. The bit-error-rate (BER) performance of the MCDM system is evaluated in simulations and indoor radio frequency (RF) experiments. We have demonstrated effectiveness of MCDM systems both in simulation and experimental results, comparing to orthogonal frequency division multiplexing (OFDM) systems. Moreover, MCDM can be further applied to higher order modulations for enabling higher data rates for RF wireless communications. 
\end{abstract}


\section{Introduction}
\label{S1}

Orthogonal frequency division multiplexing (OFDM) is a prevailing multicarrier system, which has been exploited extensively in the field of wireless communications \cite{dahlman4g,akyildiz06,nee00,wong99}. OFDM has merits of effective supporting high data rates, simultaneous multiuser transmission, and achieving maximum spectral efficiency \cite{dahlman4g}. However, OFDM symbols may suffer from harsh challenges in radio frequency (RF) channel, such as time-variant channel, multipath propagation delays, Doppler effect, and frequency selective fading \cite{akyildiz06}. 

\vspace{2pt}

Chirp signals possess characteristics of resilience and robustness to multipath propagation delays, Doppler spread, and channel fading, and their superior correlation properties, so they have been utilized broadly for numerous kinds of applications, e.g., robust underwater communication \cite{huang17}, integrated radar and communications \cite{roberton03}, frame synchronization \cite{boumard09}, Doppler compensation \cite{sharif00}, and long-range aerial communications \cite{lee15}. Moreover, chirp signals have even been incorporated as chirp spread spectrum (CSS) in IEEE 802.15.4 standard \cite{karapistoli10}. 

\vspace{2pt}

Driven by the demand of achieving higher data transmission, chirps have been further exploited in multicarrier communications. Some researchers utilized chirped OFDM for enhancing power efficiency, delay spreading, and Doppler effect \cite{dida16,barbarossa01}. Some focus on waveform designs for multiple transmitters in radar systems \cite{kim13}, orthogonal chirp signals in multiple-input multiple-output (MIMO) \cite{wang15}, and CSS for enhancing orthogonality of transmission of different users \cite{liu06,yang11,cheng15}. Several works focus on different categories of chirp bases in multicarrier systems, e.g., fractional Fourier transform (FrFT), fractional cosine transform, and Fresnel chirp transform \cite{en04,solyman12,attar17,ouyang16}. However, these chirp related transforms are of high computational overhead and complex implementations. Hence, developing an efficient and low complexity multicarrier chirp communication system is still of interests and researches for RF wireless communications.

In this paper, we propose a multicarrier communication system that applies chirp waveforms to frequency subcarriers for implementing multicarrier systems in RF wireless channel, termed as multicarrier chirp-division multiplexing (MCDM). Orthogonal linear chirp waveforms are utilized as subcarriers in MCDM systems. Then, the orthogonality of proposed chirp signals is analyzed by their cross-correlation coefficients and derived in closed-form expressions. Moreover, an orthogonal chirp transform (OCT) and inverse orthogonal chirp transform (IOCT) are developed based on proposed orthogonal chirp waveforms for implementations. In transmitter system model, we apply IOCT, so signals are transformed into time domain for transmissions. 
Receiver design includes packet synchronization, carrier frequency offset (CFO) estimation, channel estimation and symbol detection procedures. 
Packet synchronization and CFO estimation are processed in time by known pseudorandom noise (PN) training sequences. Then, we perform OCT to transform received signals into frequency domain for channel estimation and data detection. Channel information is estimated by known pilot symbols uniformly distributed among frequency subcarriers. Hence, we can estimate channel state information (CSI) every several subcarriers for updating CSI effectively. 
Then, computational complexity for our proposed detector is discussed in detail. In addition, MCDM systems are compatible with existing applications of linear chirp waveforms. The performance of MCDM systems is evaluated in the aspect of bit-error-rate (BER) in simulations and indoor RF software-defined testbeds. Therefore, we have demonstrated both in simulation and experimental results that MCDM systems can enhance performance of communications and be applied in higher order modulations for providing higher data rates for RF wireless communications. 

\vspace{1pt}

The rest of the paper is organized as follows. Section \ref{S2} introduces orthogonal chirp waveform design. System model is presented in Section \ref{S3}. Receiver design and computational complexity is elaborated in Section \ref{S4}. Simulation and experimental studies are demonstrated in Section \ref{S5}. Lastly, some concluding remarks are addressed in Section \ref{S6}.

\section{Orthogonal Chirp Waveform Design}
\label{S2}

\subsection{Chirp Waveform Design}
We consider a linear chirp waveform, which its frequency evolves linearly over time. The $k$-th chirp waveform can be represented as
\begin{align}
\psi_k(t) =  \sqrt{\frac{1}{T}} e^{j(2\pi k\Delta ft+\pi \mu t^2)},~0\leq t\leq T
\label{psi}
\end{align}
where $\Delta f$ is the frequency spacing between chirp waveforms, $\mu\triangleq\frac{B_c}{T}$ is the chirp rate, $B_c$ is the frequency span of a chirp signal, $T$ is the symbol duration, and signal energy has been normalized to 1. For the chirp rate, $\mu >$ 0 is a up-chirp waveform (frequency increasing with time), whereas $\mu <$ 0 is a down-chirp waveform (frequency decreasing with time). 

As shown in (\ref{psi}), considered chirp signals are complex consisting both in-phase and quadrature components. In addition, frequency span $B_c$ of chirp waveforms can provide an additional degree of dimension for combating multipath propagation delays and Doppler spread.

\subsection{Cross-correlation Coefficient}
The orthogonality of chirp signals is an essential metric to examine characteristics of waveforms, which can be measured by cross-correlation coefficients. The cross-correlation coefficient of the $m$-th and the $n$-th chirp waveforms is derived as
\begin{align}
\begin{aligned}
\rho_{mn}  &= \int_0^{T} \psi_{m}(t) \psi_{n}^*(t)dt\\
      &= \frac{1}{T} \int_0^{T} e^{j2\pi (m-n)\Delta ft}dt\\
      &= \delta_{mn} = \left\{\begin{tabular}{cc}
      0, &\it{m $\neq$ n}\\
      1, &\it{m = n}\\
      \end{tabular}\right.
\end{aligned}
\label{cross_corre}
\end{align}
where frequency spacing has to satisfy the criterion $\Delta f = \frac{1}{T}$ for guaranteeing the orthogonality.

From (\ref{cross_corre}), we can discover that correlation coefficients among different chirp signals are all zeros. Therefore, we have demonstrated that proposed chirp waveforms are orthogonal to each other. Hence, the principle of orthogonality of proposed chirp waveforms is guaranteed.

\subsection{Orthogonal Chirp Transform}
For representation purposes, we propose an orthogonal chirp transform (OCT) based on orthogonal chirp waveforms written as
\begin{align}
\label{OCT}
X(f) =  \sqrt{\frac{1}{T}} \int_{-\infty}^{\infty} x(t)e^{-j(2\pi ft+\pi \mu t^2)}dt
\end{align}
where $x(t)$ is an arbitrary time function and $X(f)$ is its correspondent OCT form. Hence, OCT can transform a time function into its frequency representation.

On the other hand, its correspondent inverse orthogonal chirp transform (IOCT) can be expressed as
\begin{align}
\label{IOCT}
x(t) =  \sqrt{\frac{1}{T}} \int_{-\infty}^{\infty} X(f)e^{j(2\pi ft+\pi \mu t^2)}df
\end{align}

IOCT can transform a frequency function into its time-domain form. Therefore, OCT and IOCT can transform a time function and its frequency form interchangeably, i.e., $x(t)$ and $X(f)$ constitute an OCT pair.

\section{System Model}
\label{S3}

Orthogonal chirp waveforms are further applied to frequency subcarriers for implementing MCDM systems. Then, transmitted MCDM symbol is given by
\begin{align}
\label{s_pass}
x(t) = \sqrt{E}\sum_{k=0}^{K-1} s[k]\psi_k(t) e^{j2\pi f_c t},~0\leq t\leq T
\end{align}
where $E$ is signal energy, $s[k]$ is the $k$-th transmitted symbol, $\psi_k(t)$ is the $k$-th subcarrier continuous-time function, $f_c$ is the carrier frequency, $K$ is the number of frequency subcarriers, and $T$ is the symbol duration. $s[k]$ can be a symbol of any modulations, e.g., QPSK, 16-QAM, and symbol energy has been normalized to 1.

\begin{figure}
 \centering
  \includegraphics[width=0.85\textwidth]{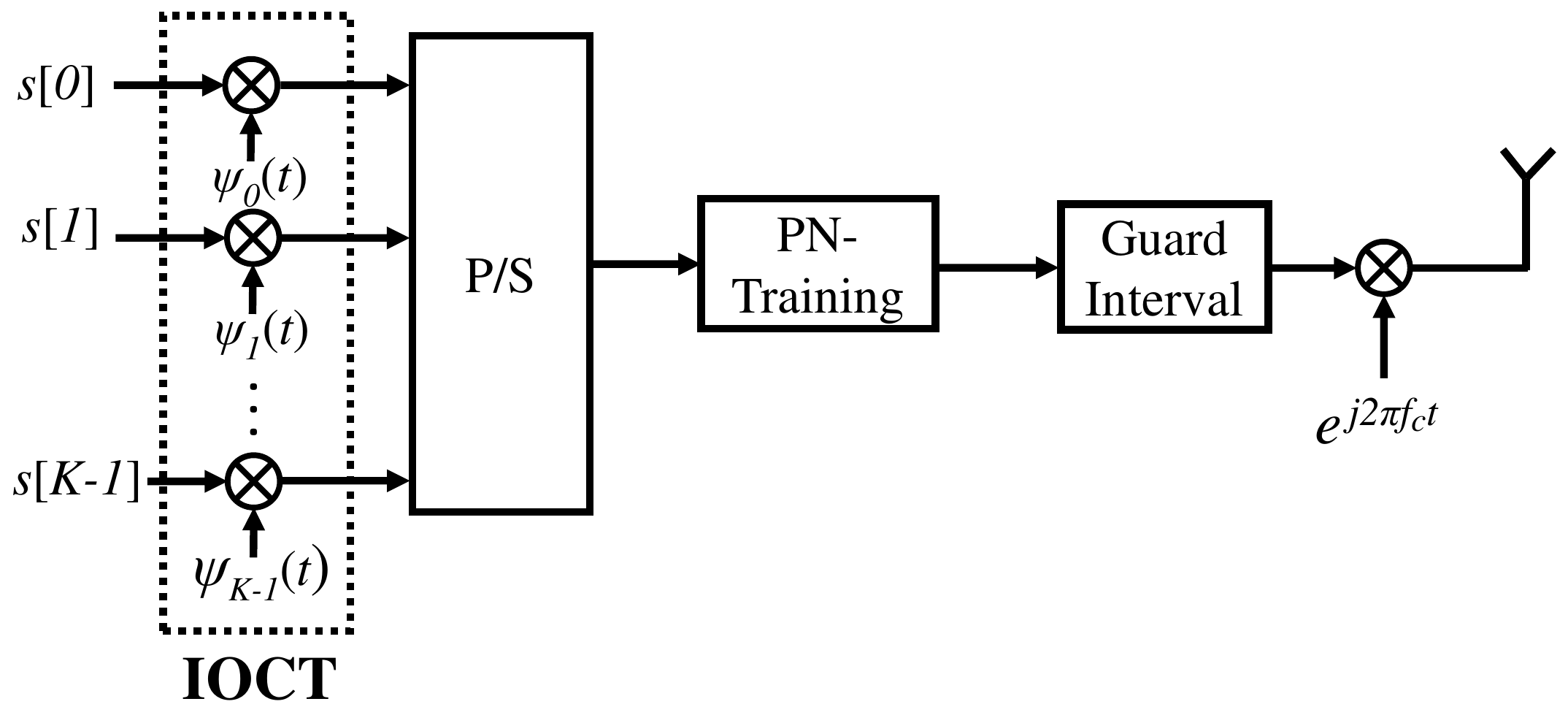} 
  \caption{A MCDM transmitter system model.}
  \label{MCDM_tx}
\end{figure}

Transmitted signals in (\ref{s_pass}) can be alternatively represented as the summed output of IOCT and after carrier frequency modulation as shown in Fig. \ref{MCDM_tx}, where transmitted symbols are prepared in frequency, and then applying IOCT to transform signals into time domain for transmissions.

In channel model, paths are assumed to be independent, multipath, and time-variant written as
\begin{align}
h(t) \overset{\triangle}{=} \sum_{m=0}^{M-1} h_m(t) \delta(t - \tau_m(t))
\label{eq:3}
\end{align}
where $M$ is the number of resolvable paths, $h_m(t)$ is the $m$-th path's amplitude, and $\tau_m(t)$ is the $m$-th path's delay. 

We assume that each path's amplitude and delay do not change over time in the duration of a MCDM symbol. Then, received baseband signals after carrier demodulation is given by
\begin{align}
r(t) = \sum_{k=0}^{K-1}\sum_{m=0}^{M-1} {\tilde h}_m s[k]\psi_k(t- \tau_m) + n(t)
\label{eq:4}
\end{align}
where ${\tilde h}_m \triangleq \sqrt{E} h_m e^{-j2\pi f_c \tau_m} \in \mathbb{C}$ is the $m$-th path's energy-including channel coefficient and $n(t)$ is additive noise.

\begin{figure}
 \centering
  \includegraphics[width=0.85\textwidth]{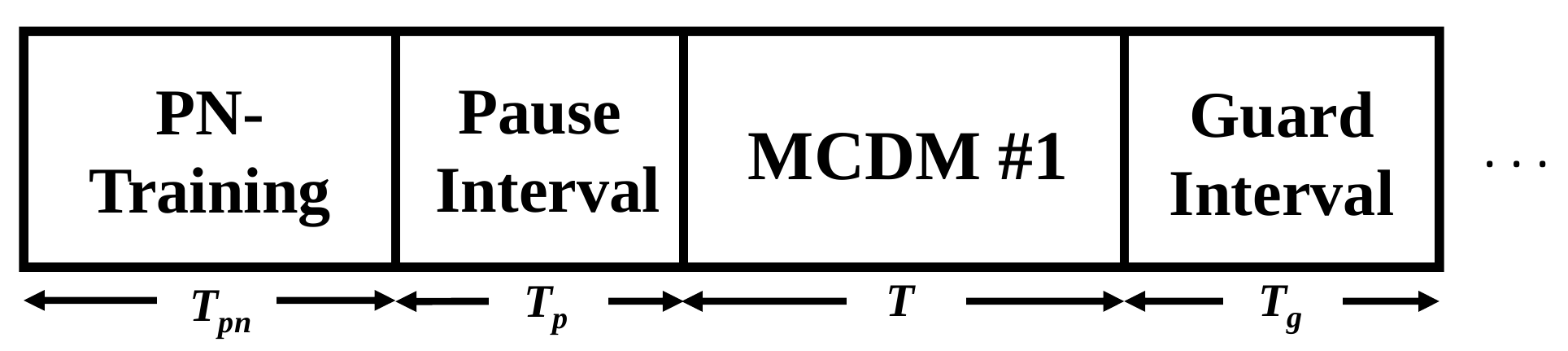} 
  \caption{MCDM transmitted packet structure.}
  \label{packet}
\end{figure}

The transmitted packet structure of a MCDM system is depicted in Fig. \ref{packet}. A PN-training block containing of antipodal sequences $\in \{\pm 1\}^{N_{pn}}$ precedes the transmission for packet synchronization. A MCDM symbol consists of multicarrier chirp signals in the duration $T$, and a pause interval of a time duration $T_p$ containing of all zeros is inserted between a PN-training block and the first MCDM symbol. On the other hand, we adopt zero-padded (ZP) technique to protect MCDM symbols from multipath propagation delays and inter symbol interference (ISI) in MCDM systems, so a guard interval of a time duration $T_g$ separates MCDM symbols. Therefore, the proposed multicarrier architecture is a ZP-MCDM system. 

Our problem objective is to design orthogonal chirp waveforms and utilize them as frequency subcarriers. OCT and IOCT are proposed for implementing MCDM systems. Moreover, a low complexity receiver is developed for reducing computational overhead in implementations.

\section{Receiver Design}
\label{S4}

\begin{figure}
 \centering
  \includegraphics[width=0.85\textwidth]{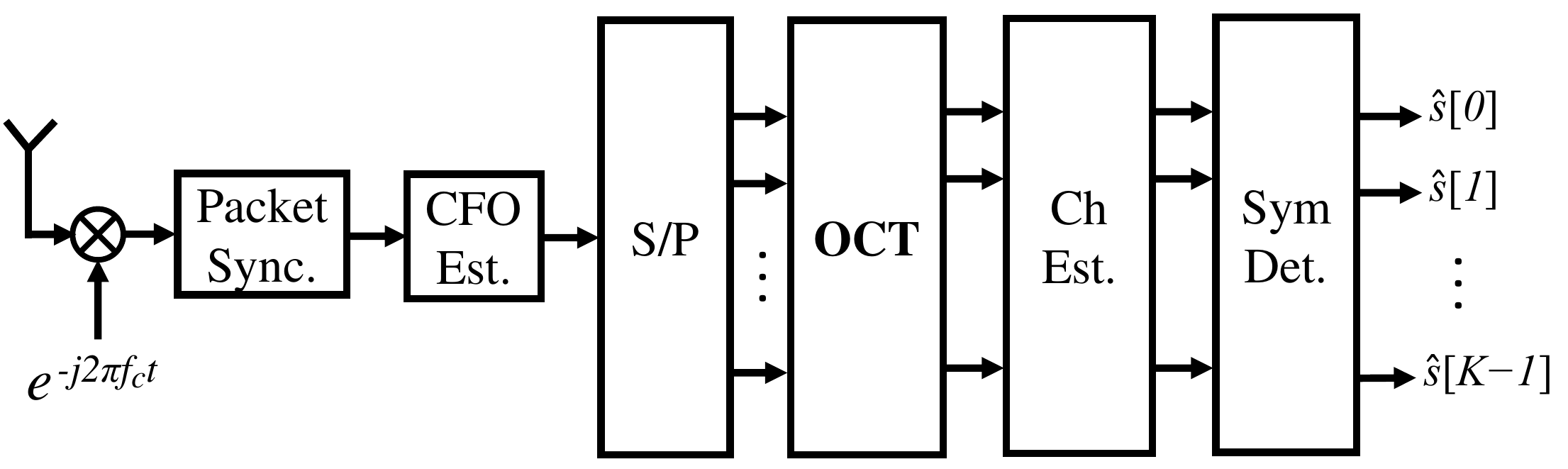} 
  \caption{MCDM system receiver design.}
  \label{MCDM_rx}
\end{figure}

The receiver design is illustrated in Fig. \ref{MCDM_rx}, including packet synchronization, CFO estimation, channel estimation, and symbol detection.

\subsection{Packet Synchronization}
\begin{figure}
 \centering
  \includegraphics[width=0.85\textwidth]{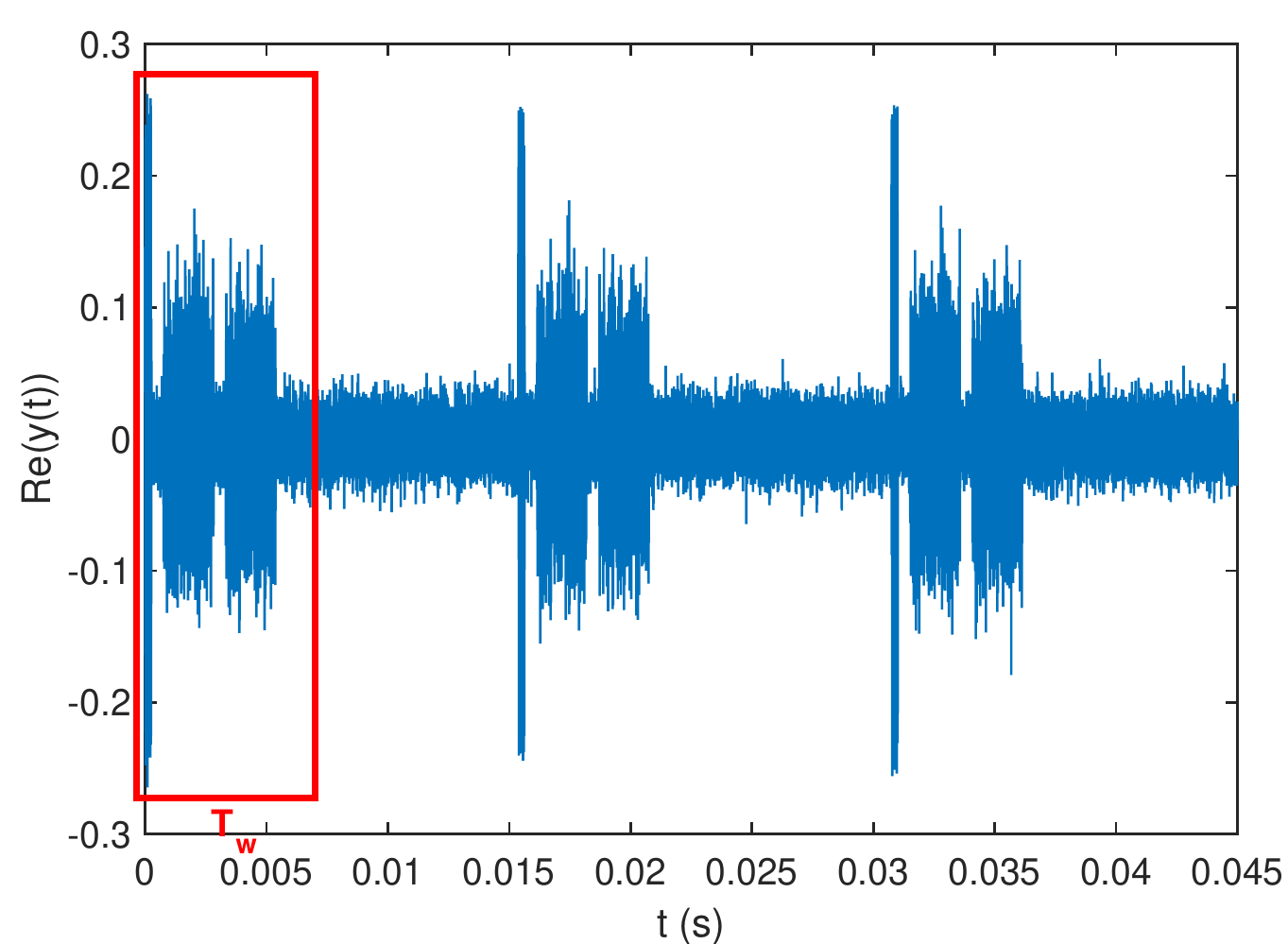} 
  \caption{Received signals observed in time.}
  \label{window}
\end{figure}

First, carrier demodulation is performed, so the following signal processing is conducted in baseband. Then, received signals are considered in a time window $T_w$ for packet synchronization as illustrated in Fig. \ref{window}. 

Packet synchronization is processed in time by a priori information of PN-training sequences of a time duration $T_{pn}$ consisting of antipodal bits. The correlation of transmitted PN-training sequences and received signals is given by
\begin{align}
\label{auto}
R(\tau) = \int_0^{T_{pn}} s_{pn}(t)r^*(t+\tau)dt
\end{align}
where $s_{pn}(t)$ is the transmitted PN-training signal, $r(t+\tau)$ is the received signal shifted with $\tau$, and $\tau \in [0,~T_w]$.

Synchronization of the packet transmission happens at the maximum of the absolute value of the correlation output written as
\begin{align}
\label{k_syn}
{\hat{t}_p} = arg\ \max_{\tau \in [0,~T_w]} \left | \left | R(\tau) \right |\right |
\end{align}

\subsection{Carrier Frequency Offset}
Signals may suffer from CFO issue due to disturbance of RF channel, Doppler shift, or oscillators' frequencies' mismatch between transmitters and receivers, which demands to be estimated and compensated for reducing errors in data detection. 

An efficient CFO estimation time-domain approach can be processed by two identical sets of PN-training sequences \cite{sourour04}, where the only difference between two sets of PN-training sequences is their time indexes. Hence, we can measure their phase difference to estimate the CFO component. The carrier frequency offset estimator is represented as
\begin{align}
\label{cfo}
{\widehat{\Delta f}} = \frac{1}{\pi T_{pn}} \angle \int_{0}^{\frac{T_{pn}}{2}} r^*_{pn}(t)r_{pn}\left(t+\frac{T_{pn}}{2}\right)dt
\end{align}
where $r_{pn}(t)$ is the received PN-training continuous-time function, and $\angle(.)$ is a phase measurement. The first set of PN-training sequences is in time range $\in [0,~\frac{T_{pn}}{2}]$, and the second set contains identical PN-training sequences, whereas shifted in time for $\frac{T_{pn}}{2}$. CFO component is averaged over the duration $\frac{T_{pn}}{2}$.


\subsection{Channel Estimation}
Next, we perform OCT to transform back into frequency domain and sample over a duration $T$ at sampling frequency $f_s$. The reason for conducting OCT is to inverse IOCT procedure in the transmitter. After OCT and sampling, the received MCDM vector $\mathbf{y}$ is written as
\begin{align}
\label{y_base}
\mathbf{y} = \mathbf{H} \mathbf{s} + \mathbf{n}~\in \mathbb{C}^{K}
\end{align}
where $\mathbf{s} \in \mathbb{C}^{K}$ is the sampled symbol vector, $\mathbf{n}\in \mathbb{C}^{K}$ is the noise vector, $\mathbf{H}$ is the channel matrix
\begin{align}
{\mathbf H} \triangleq \left[\begin{array}{ccccc}
       h_0    & 0      & \cdots  & 0       & 0\\
       0      & h_1    & \cdots  & 0       & 0\\
       \vdots &\vdots  & \ddots  & \vdots  & \vdots\\
       0      & 0      & \cdots  & h_{K-2} & 0\\
       0      & 0      & \cdots  & 0       & h_{K-1}
     \end{array}\right]\in \mathbb{C}^{K \times K}
\label{eq:6}
\end{align}
where $h_k$ is the channel coefficient of the $k$-th subcarrier, and $K$ is the total number of subcarriers. We assume that bandwidth of a subcarrier is smaller than channel coherence bandwidth \cite{coulson01}, so the channel fading can be viewed as flat within the interval of a subcarrier. Based on these assumptions, the channel coefficient can be estimated as a complex number, so multipath channel estimation is not of need, which the computational complexity can be reduced significantly.

\begin{figure}
 \centering
  \includegraphics[width=0.85\textwidth]{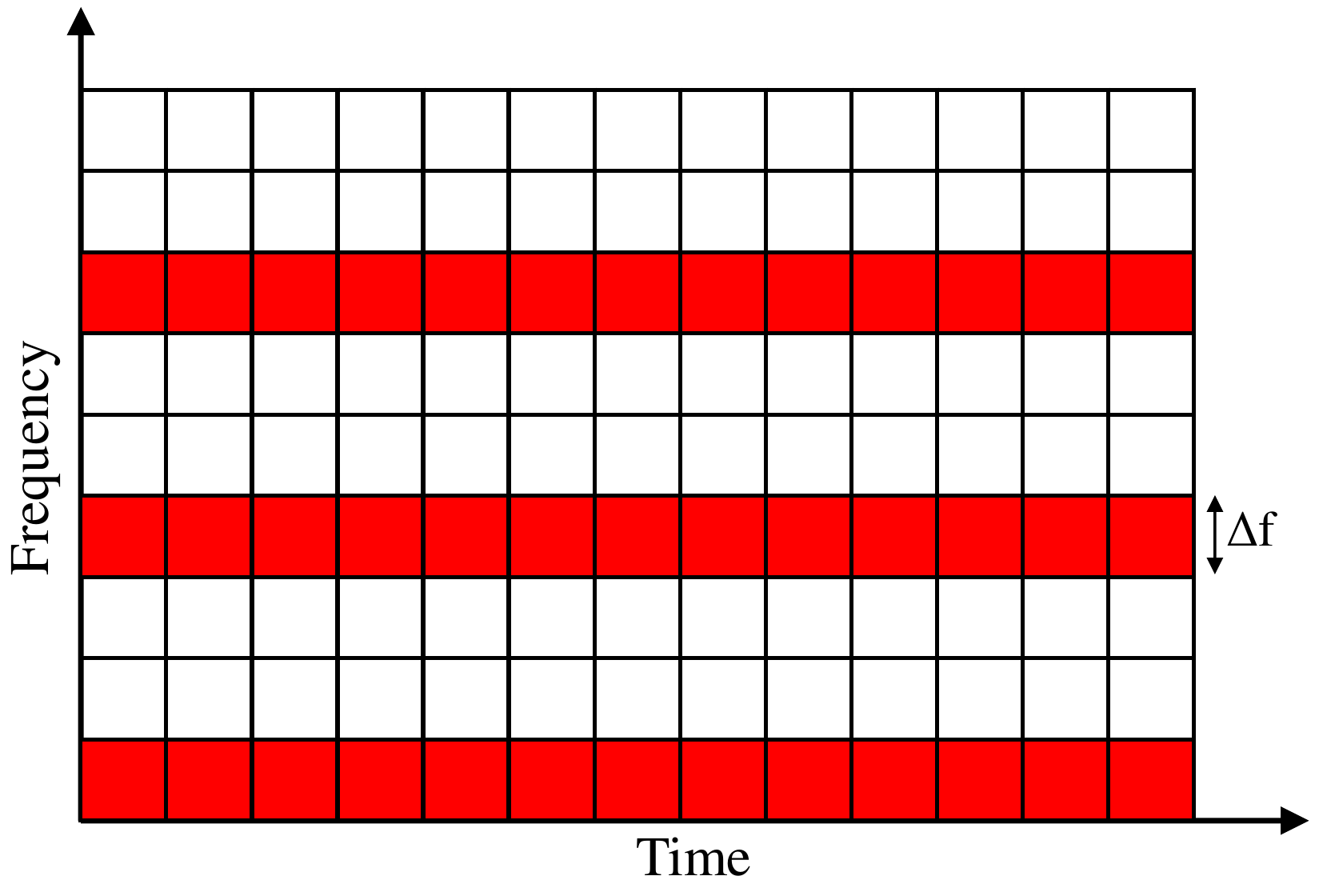} 
  \caption{Comb-type pilot subcarrier allocation.}
  \label{sub_pilot}
\end{figure}

For pilot allocation, comb-type pilot subcarrier allocation is adopted in a MCDM system as shown in Fig. \ref{sub_pilot}, where pilot symbols in red cells are spread uniformly among subcarriers in frequency \cite{shen06} and subcarrier allocation is fixed over time. Pilot symbols are antipodal binary bits $s_{p,k_p} \in \{\pm 1\}^{K_p-1}_{k_p=0}$, which are evenly distributed at locations $k$ = 0, $L$,~ $2L$,~\ldots~,$(K_p-1)L$, where $L \triangleq \frac{K}{K_p}$. Therefore, $K$ subcarriers are divided into $K_p$ groups, each group containing $L$ subcarriers, so channel states are estimated every $L$ subcarriers for effective updating channel information.
 
Then, the received pilot vector $\mathbf{y}_p$ is represented as
\begin{align}
\label{y_pilot}
\mathbf{y}_{\it p} = \mathbf{S}_{\it p} \mathbf{h}_{\it p} + \mathbf{n}_{\it p}~\in \mathbb{C}^{K_p}
\end{align}
where $\mathbf{h}_{\it p} = [h_{p,0},\ h_{p,1},\ \cdots,\ h_{p,K_p-1}]^T \in \mathbb{C}^{K_p }$ is the pilot channel vector, $\mathbf{n}_{\it p} \in \mathbb{C}^{K_p }$ is the pilot noise vector, and $\mathbf{S}_{\it p}$ is the pilot matrix
\begin{align}
{\mathbf S_p} \triangleq \left[\begin{array}{ccccc}
       s_{p,0} &0       &\cdots  &0\\
       0       &s_{p,1}&\cdots   &0\\
       \vdots  &\vdots  &\ddots  &\vdots\\
       0       &0       &\cdots  &s_{p,K_p-1}
     \end{array}\right]~\in \mathbb{C}^{K_p \times K_p}
\label{s_p}
\end{align}
where $s_{p,k_p}$ is the $k_p$-th pilot symbol.

Furthermore, channel estimation for pilot symbols can be solved by the least-squares (LS) problem
\begin{align}
\label{pro_che}
\mathbf{\widehat{h}_{\it p}} &= arg\ \min_{\mathbf{h_p} \in \mathbb{C}^{K_p }} \left|\left|\mathbf{y}_{\it p} - \mathbf{S}_{\it p} \mathbf{h}_{\it p}\right|\right|^{\rm 2}_{\rm 2}
\end{align}
where $\left|\left|\cdot\right|\right|_{\rm 2}$ denotes the Euclidean norm. 

Given pilot matrix is orthogonal and energy of pilot symbol is normalized to 1, the channel coefficient of the $k_p$-th pilot subcarrier can be estimated as
\begin{align}
\label{h_LS}
{\widehat{h}_{\it p,k_p}} &= s_{p,k_p} y_{p,k_p}
\end{align} 
where $y_{p,k_p}$ is the $k_p$-th received pilot symbol. Under additive white Gaussian noise (AWGN) condition, the channel estimate is maximum-likelihood (ML) optimal \cite{Proakis07}.

Channel of each subcarrier is assumed to be narrowband and CSI changes linearly between adjacent subcarriers. Based on the estimated channel coefficients for pilot subcarriers, we therefore operate linear interpolation of channel estimation for subcarriers $k$ at non-pilot locations $k_p L < k < (k_p+1)L$ represented as
\begin{align}
\label{h_inter}
\widehat{h}_k = \left(1-\frac{l}{L}\right)\widehat{h}_{p,k_p} + \frac{l}{L}\widehat{h}_{p,k_p+1},~k= 0,\cdots, K-1
\end{align}
where $\widehat{h}_{p,k_p}$ is the estimated channel coefficient of the $k_p$-th pilot subcarrier, $l \triangleq mod(k,L)$ is the residue of $k$ divided by $L$, and $k_p = 0,\cdots, K_p-1$. 

\subsection{Symbol Detection}
Traditionally, symbol is detected by minimizing the Euclidean distance between received vector $\mathbf{y}$ and channel-filtered vector $\mathbf{Hs}$ formulated as
\begin{align}
\label{prob}
\mathbf{\widehat{s}} &= arg\ \min_{\mathbf{s} \in~\mathbb{S}^{K}} \left|\left|\mathbf{y} - \mathbf{H} \mathbf{s}\right|\right|^{\rm 2}_{\rm 2}
\end{align}
where $\mathbb{S}^{K} \in \mathbb{C}^{K}$ is the alphabet set of $K$ symbols. 

Since $\mathbf{H}$ is a diagonal matrix, symbol detection can be conducted in subcarrier-wise manner. Then, the $k$-th symbol is detected as
\begin{align}
\label{s_k}
\widehat{s}_k &= arg\ \min_{s \in~\mathbb{S}} \left|\left |y_k - h_k s_k\right|\right|^{\rm 2}, ~ k = 0,\cdots,K-1
\end{align}
where $\mathbb{S} \in \mathbb{C}$ is the alphabet set of a symbol.


\subsection{Computational Complexity}

The complexity of the minimum distance detector and the proposed detector are discussed in detail. Computational complexity for minimum distance detector: multiplication is $O(K^2D)$ and addition is $O(K^2D)$, where $D$ is the alphabet size of the symbol modulation. On the other hand, the computation for our detection approach: multiplication is $O(KD)$ and addition is $O(KD)$. As demonstrated in the comparison, the computational complexity of proposed detector design is only linear with number of subcarriers $K$ instead of $K^2$ in the minimum distance detector. Moreover, in our approach, symbols can be even detected simultaneously among subcarriers for reducing processing time. Hence, our proposed detector can save computational complexity dramatically in implementations.

\section{Simulation and Experimental Studies}
\label{S5}

\subsection{Simulation Studies}
\begin{figure}
 \centering
  \includegraphics[width=0.85\textwidth]{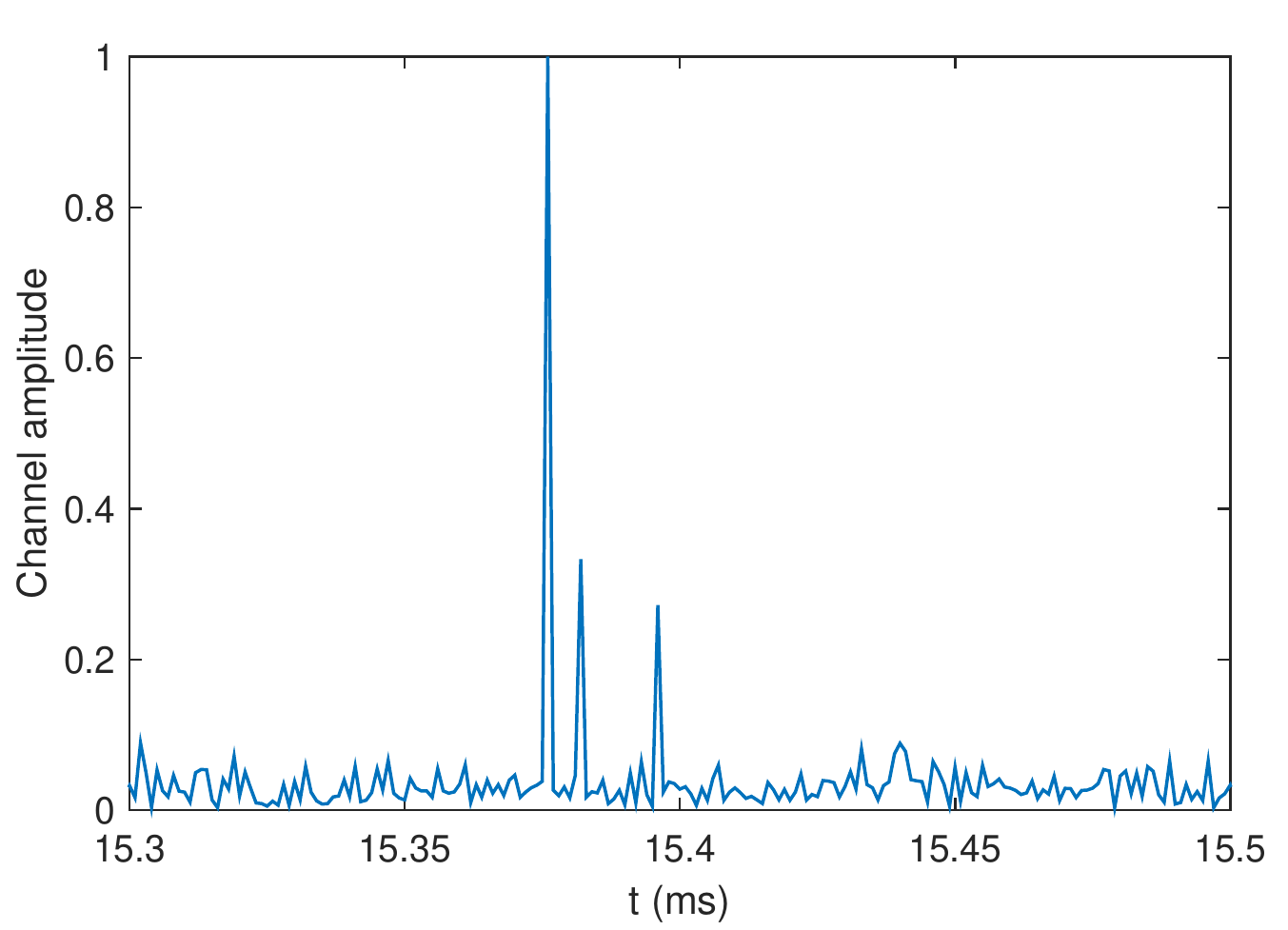} 
  \caption{Normalized channel response in simulations.}
  \label{ch_Sim}
\end{figure}

In simulations, MCDM systems are simulated in RF wireless emulators. The amplitude of channel response is modeled as Rayleigh distribution and noise is white Gaussian. Normalized channel response is shown in Fig. \ref{ch_Sim}. We consider several modulations, including binary phase shift keying (BPSK), quadrature PSK (QPSK), 8-PSK, 16 quadrature amplitude modulation (16-QAM), and 32-QAM. Moreover, bit mapping for symbols is gray coded for reducing detection errors between adjacent symbols. Total number of subcarriers $K$ is $1024$, including pilot subcarriers $K_p$ = $256$, null carriers $K_n$ = $56$, and data subcarrier $K_s$ = $712$. Carrier frequency is $f_c$ = $2.42$ GHz. The frequency spacing between subcarriers $\Delta f$ = $488$ Hz. The MCDM symbol period is $T$ = $2.05$ ms, the preamble duration $T_{pn}$ = $0.26$ ms, the pause interval between preambles and a MCDM symbol $T_p$ = $0.51$ ms, and the guard interval between MCDM symbols is $T_g$ = $0.51$ ms. Linear up-chirps are applied, chirp rate $\mu$ chosen as $2.38 \cdot 10^5$ Hz/s. Available bandwidth for the MCDM system is $500.50$ kHz. 

\begin{figure}
 \centering
  \includegraphics[width=0.85\textwidth]{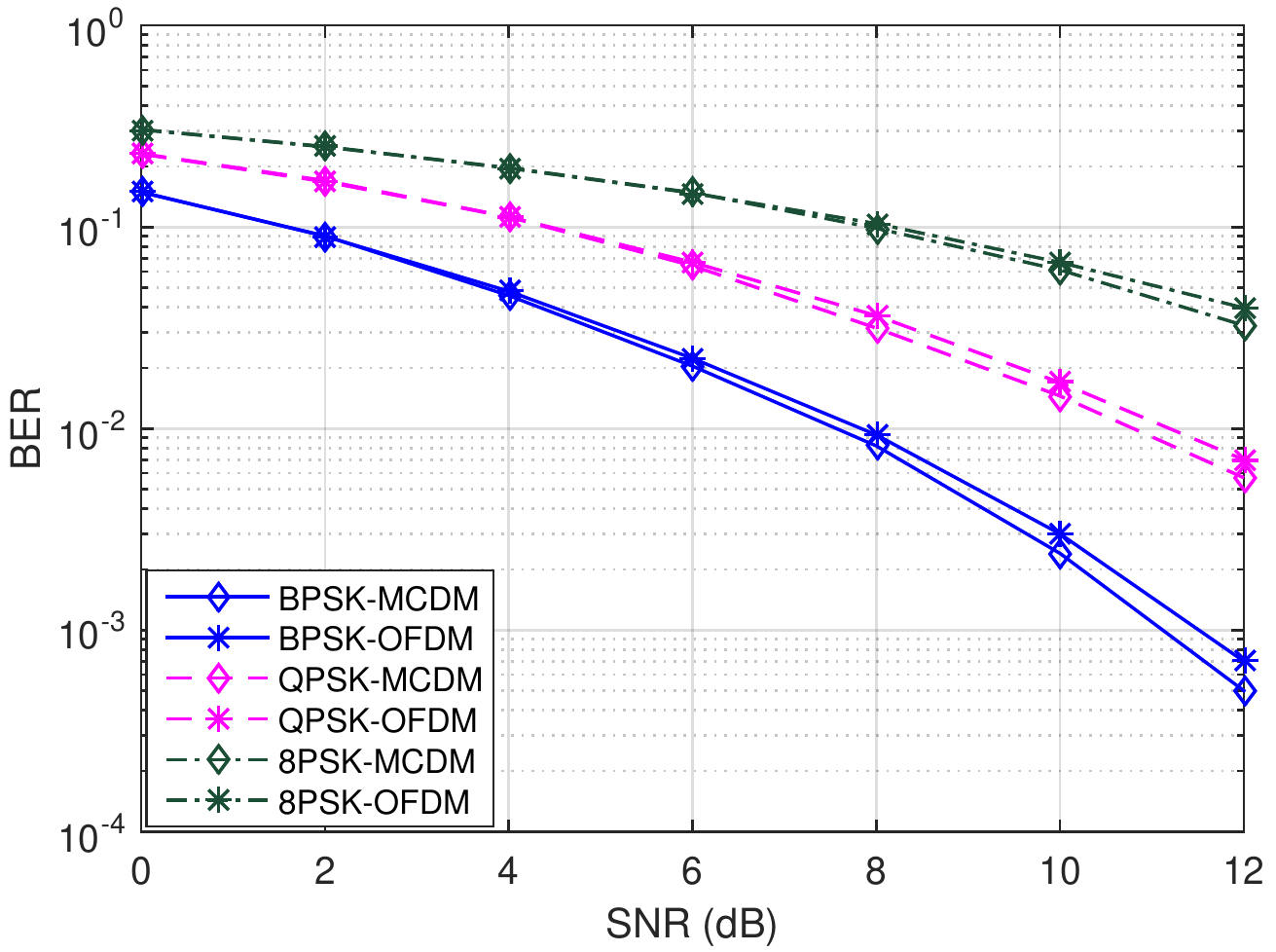} 
  \caption{Performance of PSK-MCDM in simulations.}
  \label{PSK_Sim}
\end{figure}

\begin{figure}
 \centering
  \includegraphics[width=0.85\textwidth]{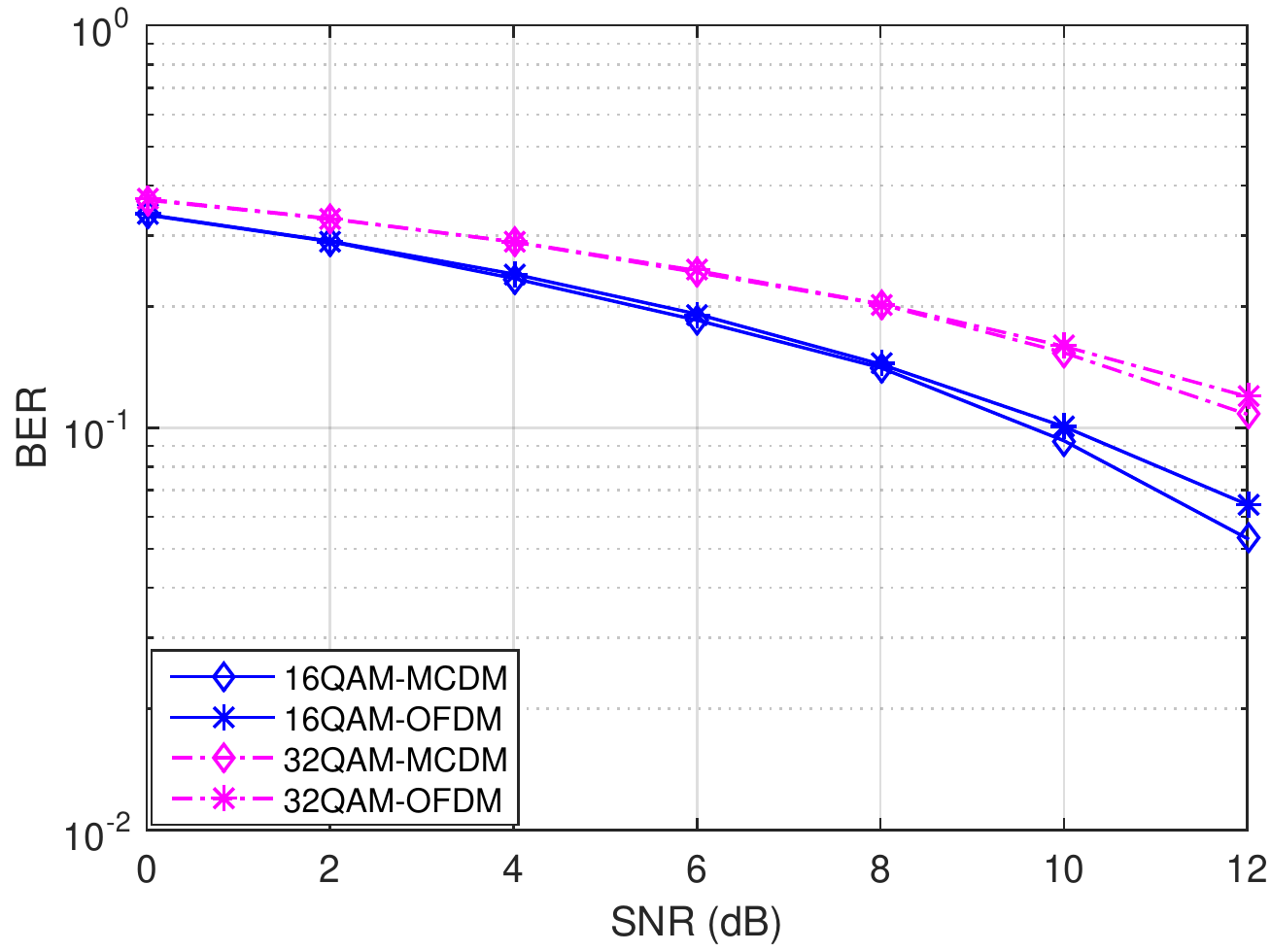} 
  \caption{Performance of QAM-MCDM in simulations.}
  \label{QAM_Sim}
\end{figure}

Fig. \ref{PSK_Sim} shows BER of various PSK modulations over SNR values. For BPSK at SNR = $12$ dB, BER of MCDM is enhanced for $1.46$ dB than that of OFDM. For QPSK at SNR = $10$ dB, BER of MCDM is improved for $0.67$ dB than BER of OFDM. In MCDM systems, bit rate of 8-PSK can achieve up to $834.38$ kbit/s (kbps), while the BER of 8-PSK is $3.25 \times 10^{-2}$ at SNR = $12$ dB, which is $7.56$ dB higher than that of QPSK. Therefore, there is a trade-off between bit rate and BER for designing communication systems. 
 
Fig. \ref{QAM_Sim} illustrates BER performance of various QAM modulations versus SNR. For 16-QAM at SNR = $12$ dB, BER of MCDM is superior for $0.85$ dB than BER of OFDM. Comparing 32-QAM to 16-QAM at SNR = $12$ dB, BER of MCDM increases $3.09$ dB, while bit rate is $25\%$ higher. We have validated in simulations that BER performance can be enhanced by MCDM systems and higher order modulations can be exploited for enabling higher data rates. Effective bit rate for 32-QAM can achieve up to:
\begin{align}
\label{tx_rate}
\frac{K_s}{T+T_g}\cdot \log_2 32 = 1.39 ~M bps
\end{align}

\subsection{Experimental Studies}
\begin{figure}
 \centering
  \includegraphics[width=0.83\textwidth,trim={0 2.4cm 0 0.2cm},clip]{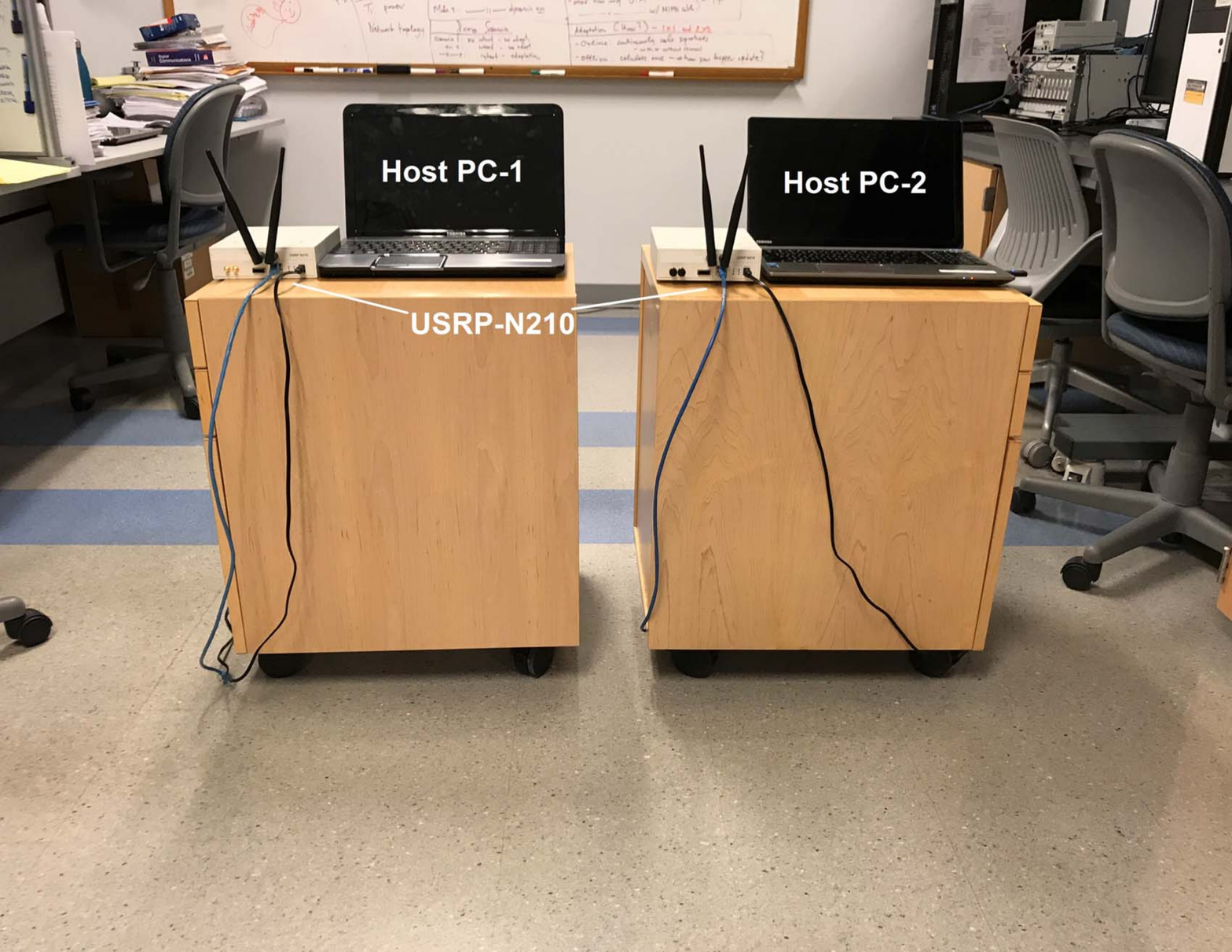} 
  \caption{Indoor SDR testbed setup for a MCDM system.}
  \label{RF_testbed}
\end{figure}

\begin{figure}
 \centering
  \includegraphics[width=0.85\textwidth]{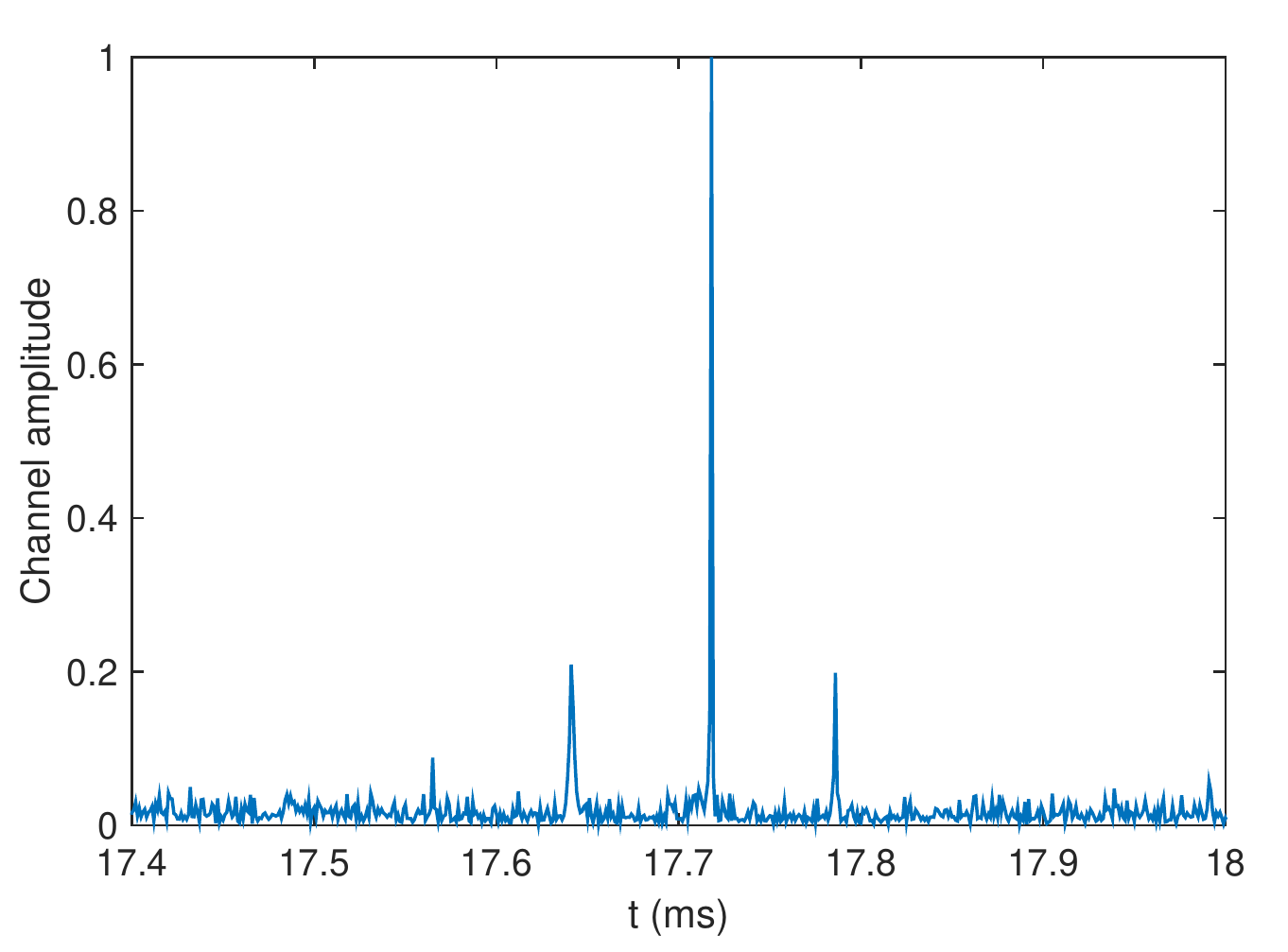} 
  \caption{Normalized channel response in experiments.}
  \label{ch_Exp}
\end{figure}

Performance of MCDM systems are evaluated in an indoor software-defined radio (SDR) testbeds depicted in Fig. \ref{RF_testbed}. Two USRP-N210s are interfaced with GNU Radio operated by two host personal computers (PCs), where one setup functions as a transmitter and the other one as a receiver deployed in two adjacent rooms, respectively. The transmission operates at $f_c$ = $2.47$ GHz, and sampling frequency $f_s$ = $1$ MHz. The considered modulations include BPSK, QPSK, 8-PSK, 16-QAM, and 32-QAM. Moreover, gray code is also utilized in bit mapping for reducing detection errors. Total number of subcarriers $K$ = $1024$, including pilot subcarriers $K_p$ = $256$, null carriers $K_n$ = $56$, and data subcarrier $K_s$ = $712$. The frequency spacing between subcarriers $\Delta f$ = $488$ Hz. The MCDM symbol period, the PN-training duration, the pause interval between PN-training and a MCDM symbol, and the guard interval between MCDM symbols are $T$ = $2.05$ ms, $T_{pn}$ = $1.02$ ms, $T_p$ = $0.51$ ms, and $T_g$ = $0.51$ ms, respectively. Linear up-chirps are adopted, chirp rate $\mu$ chosen as $2.44 \cdot 10^5$ Hz/s. Assigned bandwidth for MCDM systems is $500.50$ kHz. The normalized channel response is illustrated in Fig. \ref{ch_Exp}, consisting of 3 resolvable paths. The experimental results are averaged over $3000$ recorded packets. 

\begin{figure}
  \centering
  \includegraphics[width=0.85\textwidth]{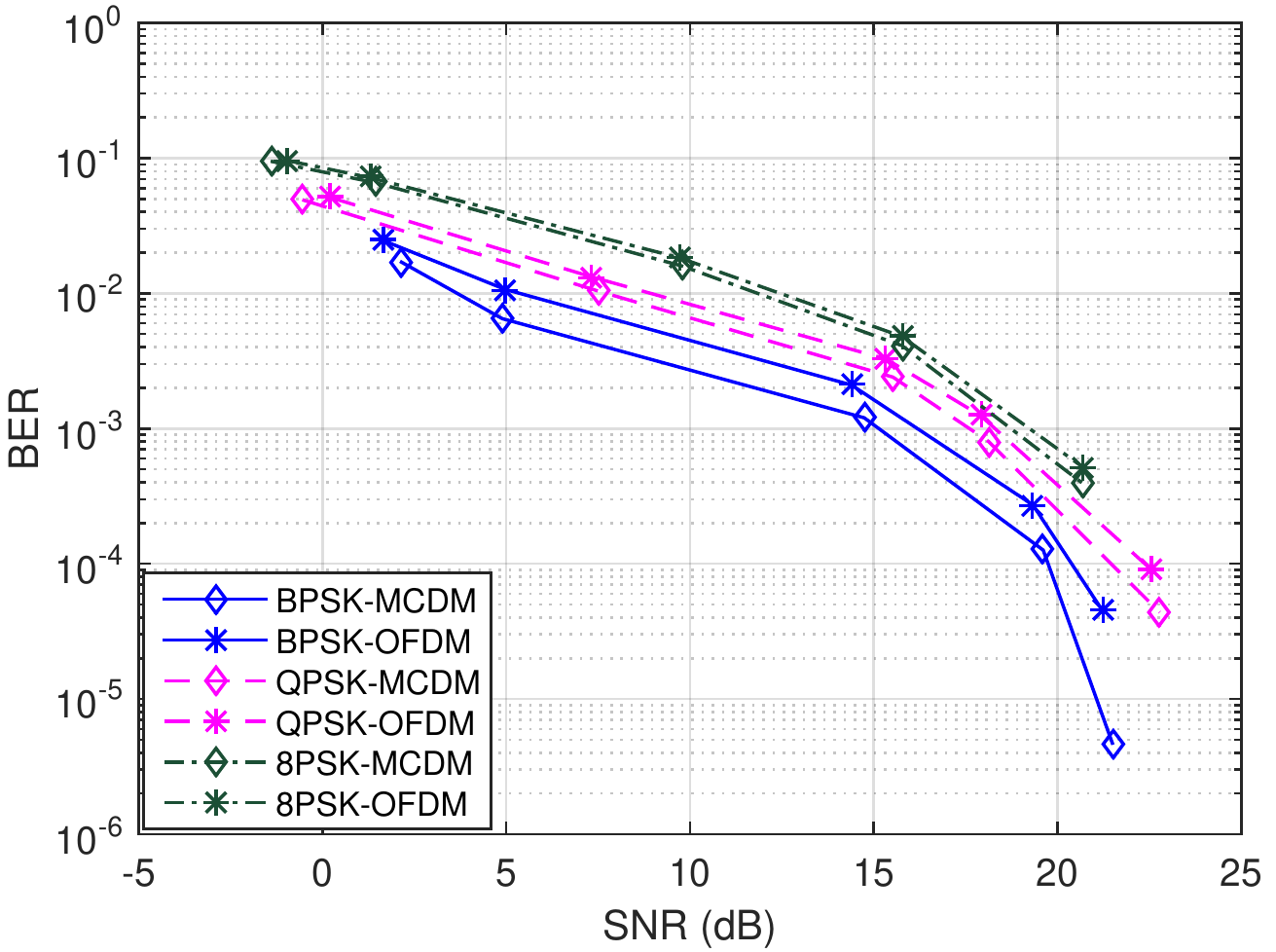} 
  \caption{Performance of PSK-MCDM in experiments.}
  \label{PSK_Exp}
\end{figure}

Fig. \ref{PSK_Exp} shows BER of PSK-MCDM in indoor RF experiments. The SNR here is the pre-detection SNR. For BPSK at SNR = $21.49$ dB, BER of MCDM decreases $9.64$ dB than that of OFDM. We can discover that BER improvement between MCDM and OFDM is the most significant for SNR $> 20$ dB. This can be explained that in high SNR region, noise is less influential, while multipath still exist to deteriorate BER performance. Then, utilization of chirp signals in MCDM systems can enhance their resistance to multipath. For MCDM systems at SNR = $18.12$ dB, BER of QPSK is $8.03 \cdot 10^{-4}$ and that of 8-PSK is $1.45 \cdot 10^{-3}$. The reason for BER difference is that symbol energy is fixed among modulations, so the higher order the modulation, the closer the Euclidean distance between adjacent constellation symbols, which increases BER values, whereas higher order modulations can instead provide higher data rates.

\begin{figure}
  \centering
  \includegraphics[width=0.85\textwidth]{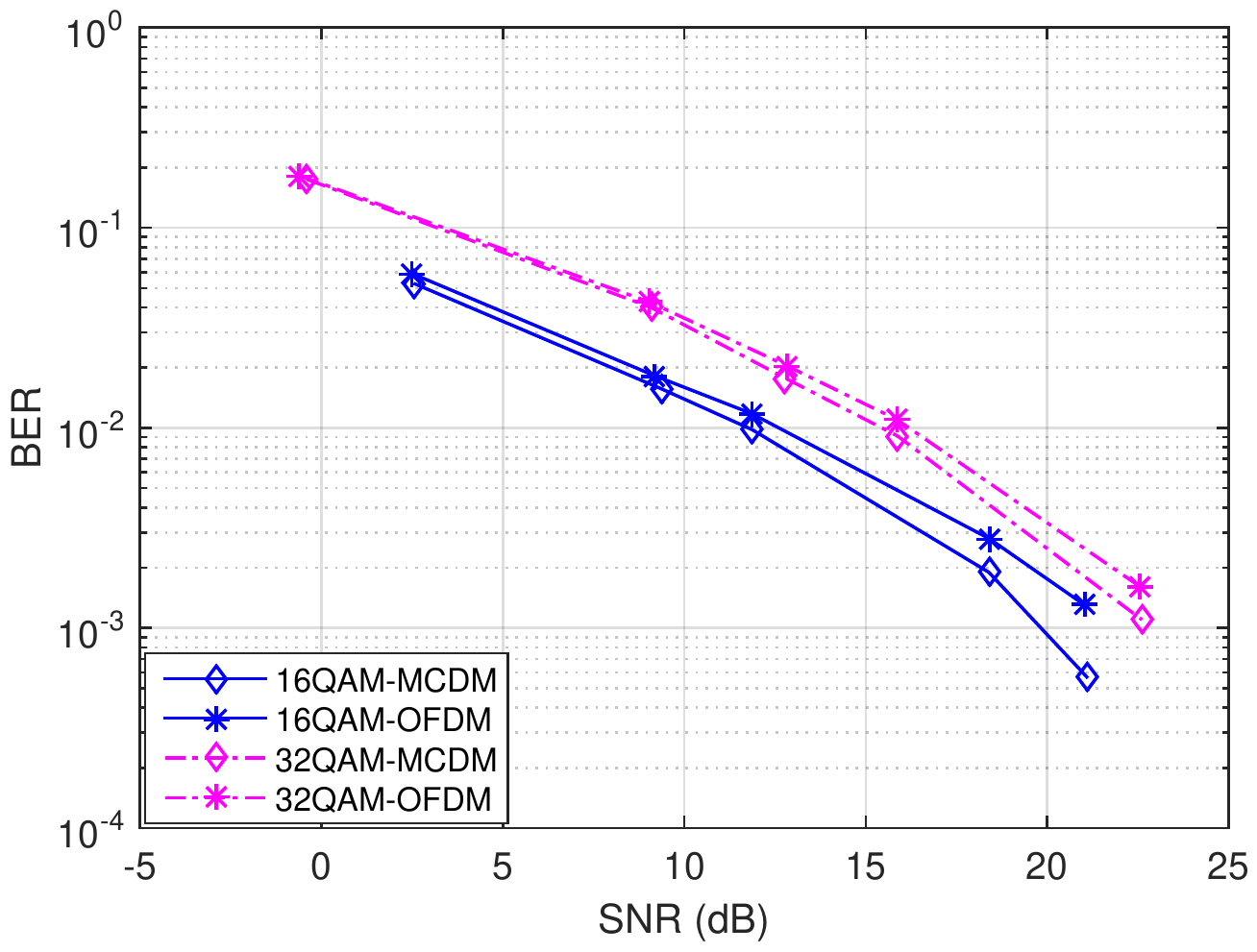} 
  \caption{Performance of QAM-MCDM in experiments.}
  \label{QAM_Exp}
\end{figure}

Fig. \ref{QAM_Exp} demonstrates BER performance of QAM-MCDM over pre-detection SNR in indoor RF experiments. For 16-QAM at SNR = $21.04$ dB, MCDM enhances BER for $3.58$ dB than that of OFDM, which is better than improvement of 8-PSK between MCDM and OFDM. This can be explained that QAM is not a phase modulation, so it can be more robust to phase errors introduced in transceivers or RF fading channel. In 32-QAM, BER of MCDM at SNR = $22.58$ dB is enhanced for $1.63$ dB than that of OFDM, which is less than BER improvement of 16-QAM due to the closer Euclidean distance between constellation symbols. At SNR = $22.58$ dB, 32-QAM MCDM system can obtain BER = $1.12 \cdot 10^{-3}$ and bit rate can achieve up to: $K_s/(T+T_g)\cdot \log_2 32$ = $1.39$ M bit/s.

\section{Conclusion}
\label{S6}

In this paper, we have proposed multicarrier chirp-division multiplexing architecture for RF wireless communications. Orthogonality of linear chirp signals is analyzed by their cross-correlation coefficients. Moreover, based on orthogonal chirp waveforms, OCT and IOCT are developed for implementations of MCDM systems. In receiver design, we utilize known PN-training symbols for packet synchronization and CFO estimation. Channel estimation is processed by the comb-type pilot subcarrier allocation, which pilot symbols are uniformly distributed among subcarriers, so CSI is updated every several subcarriers. In addition, a low complexity detector is designed to perform symbol detections. Computational complexity is analyzed for our proposed detector explicitly. Then, BER performance is evaluated in indoor RF testbeds. Experimental results demonstrate that performance of communications is enhanced by MCDM systems, and they can support higher data rate transmissions. Therefore, MCDM systems can enhance performance of high-rate communications in RF fading channel.


\linespread{1.003}

\bibliographystyle{IEEEtran}
\bibliography{ref}

\end{document}